\documentclass[11pt,a4paper]{article}
\usepackage{graphicx}
\usepackage{bm}
\usepackage{color}

\begin{document}

\begin{center}
{\huge \bf Teleparallelism}
\vskip 0.5cm
{\large \bf A New Way to Think the Gravitational Interaction}$^\dagger$
\end{center}
\vskip 1.0cm
\begin{quote}

At the time it celebrates one century of existence, general relativity~---~Einstein's theory for gravitation~---~is given a companion theory: the so-called teleparallel gravity, or teleparallelism for short. This new theory is fully equivalent to general relativity in what concerns physical results, but is deeply different from the conceptual point of view. Its characteristics make of teleparallel gravity an appealing theory, which provides an entirely new way to think the gravitational interaction.

\vskip 1.8cm
\flushright{\bf R. Aldrovandi and J. G. Pereira}\\
{\it Instituto de F\'{\i}sica Te\'orica \\
Universidade Estadual Paulista \\
S\~ao Paulo, Brazil} \\
%jpereira@ift.unesp.br

\end{quote}
\vfill
$^\dagger$ {\footnotesize English translation of the Portuguese version published in {\it Ci\^encia Hoje} {\bf 55} (326), 32 (2015).}

\newpage
%%%%%%%%%%%%%%%%%%%%%%%%%%%%%
\section*{Gravitation is universal}
%%%%%%%%%%%%%%%%%%%%%%%%%%%%%

One of the most intriguing properties of the gravitational interaction is its universality. This hallmark states that all particles of nature feel gravity the same, independently of their masses and of the matter they are constituted. If the initial conditions of a motion are the same, all particles will follow the same trajectory when submitted to a gravitational field. The origin of universality is related to the concept of mass. In principle, there should exist two different kinds of mass: the inertial mass $m_i$ and the gravitational mass $m_g$. The inertial mass would describe the resistance a particle shows whenever one attempts to change its state of motion, whereas the gravitational mass would describe how a particle reacts to the presence of a gravitational field. Particles with different relations $m_g/m_i$, therefore, should feel gravity differently when submitted to a given gravitational field. Up to now, however, no experiment succeeded in detecting any difference between those masses. The equality of $m_i$ and $m_g$ has then been assumed to be exact and incorporated into physics with the name Weak Equivalence Principle. This principle, in turn, gives rise to the so-called universal law of gravitation, established by Isaac Newton (1642-1726) in 1687. Although promoted to a principle, no one knows to what extent the equality between inertial and gravitational masses is in fact exact. This is actually an open question, and there are currently several projects aiming at improving the precision of the measurements of a possible difference between $m_i$ and $m_g$.

%%%%%%%%%%%%%%%%%%%%%%%%%%%%%
\section*{If there exists no force ...}

An important property of Newtonian gravity is that it is consistent with the relativity established by the Italian physicist, mathematician and astronomer Galileu Galilei (1564-1642), according to which the velocities are additives. For example, if a car is moving with a velocity of 50 kilometers per hour and encounters another car traveling in the opposite direction with a velocity of 30 kilometers per hour, the relative velocity between the cars will be 80 kilometers per hour. On the other hand, the electromagnetic theory, fully established by mid nineteen century, was clearly inconsistent with Galilei relativity. Further theoretical and experimental studies provided compelling indications that a new relativity that would comply with the electromagnetic theory was necessary. These studies triggered a search for this theory, which culminated in the advent of special relativity in 1905. In this theory, the relation between velocities was no longer additive, but satisfied a more complicated rule in such a way that the velocity of light turned out to be the same in all frames~---~that is to say, a constant of nature.

A new problem thus emerged: because it was not consistent with special relativity, Newtonian gravity was conceptually problematic. The basic task was then to construct a new gravitational theory that would be consistent with special relativity. In other words, to construct a relativistic theory for gravitation. In 2015, roughly ten years after the advent of special relativity, Albert Einstein (1879-1955) unveiled a new gravitational theory, which he called general relativity. For weak gravitational fields and small velocities, this theory yields the same results of Newtonian gravity. In a general case, however, it gives rise to new gravitational phenomena. Furthermore, it describes gravitation in a complete unusual way: whereas in Newtonian gravity the interaction is described by a force, in general relativity the concept of gravitational force does not exist at all.
%%%%%%%%%%%%%%%%%%%
\begin{figure}[htb]
\begin{center}
\scalebox{0.61}{\includegraphics{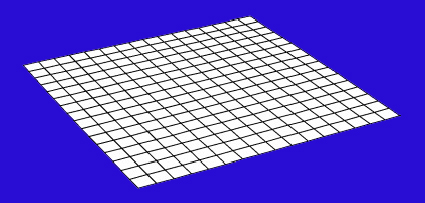}}
\caption{\it Pictorial representation of a flat two-dimensional space, that is, of a space with vanishing curvature.}
\label{fig}
\end{center}
\end{figure}
%%%%%%%%%%%%%%%%%%%

%%%%%%%%%%%%%%%%%%%%%%%%%%%%%
\section*{General relativity: curved spaces}

The question then arises: if there exists no force, how does general relativity describe the gravitational interaction? The answer is surprisingly simple. The presence of a massive body produces a gravitational field in its neighbourhood~---~which manifests itself by producing a curvature in space. For example, the Sun produces a curvature in its neighbourhood, whose intensity becomes smaller and smaller as one goes far away, vanishing at infinity~---~where the gravitational field is supposed to vanish. In absence of gravitation, therefore, space must have vanishing curvature, that is to say, must be flat. In Figure 1 a flat space, reduced to two dimensions to make visualization easier, is depicted. If an object is moving freely in such flat space, it will follow a straight line. On the other hand, if an object is moving freely in the neighbourhood of the Sun, its trajectory will naturally follow the curvature of space. 
%%%%%%%%%%%%%%%%%%%
\begin{figure}[http]
\begin{center}
\scalebox{0.5}{\includegraphics{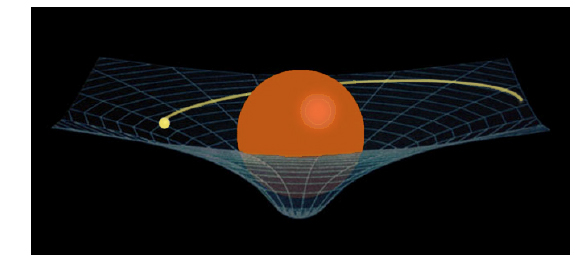}}
\caption{\it The gravitational field of the Sun produces a curvature in space, and any free object in its neighbourhood will travel along trajectories that follow the space curvature.}
\label{fig0}
\end{center}
\end{figure}
%%%%%%%%%%%%%%%%%%%
These free trajectories~---~that is, of objects without self-propulsion~---~which follow the curvature of space, are called geodesics. Considering again a two dimensional space, but keeping the Sun pictorially represented in three dimensions, Figure 2 depicts both the curvature produced by the Sun and a trajectory followed by a free object traveling in its neighbourhood. We see in this way that in general relativity the responsibility of describing the gravitational interaction is transferred to the geometry of space. There is no a gravitational force: the trajectories of free objects in a gravitational field are determined by the curvature of space. One then says the gravitational interaction is {\em geometrized}.

Depending on the initial conditions, a free object may follow a trajectory describing an open orbit, or may be trapped in the form of a closed orbit. Figure 3 shows an open hyperbolic orbit `a', and two closed orbits, an elliptic orbit `e' and a circular orbit `c'. It is important to remark that, upon transferring to the space the task of describing the gravitational interaction, the universality of gravitation is automatically incorporated in the theory: all free objects, independently of their masses and constitutions, will travel along a trajectory that follows the curvature of space.

%%%%%%%%%%%%%%%%%%%
\begin{figure}[http]
\begin{center}
\scalebox{0.7}{\includegraphics{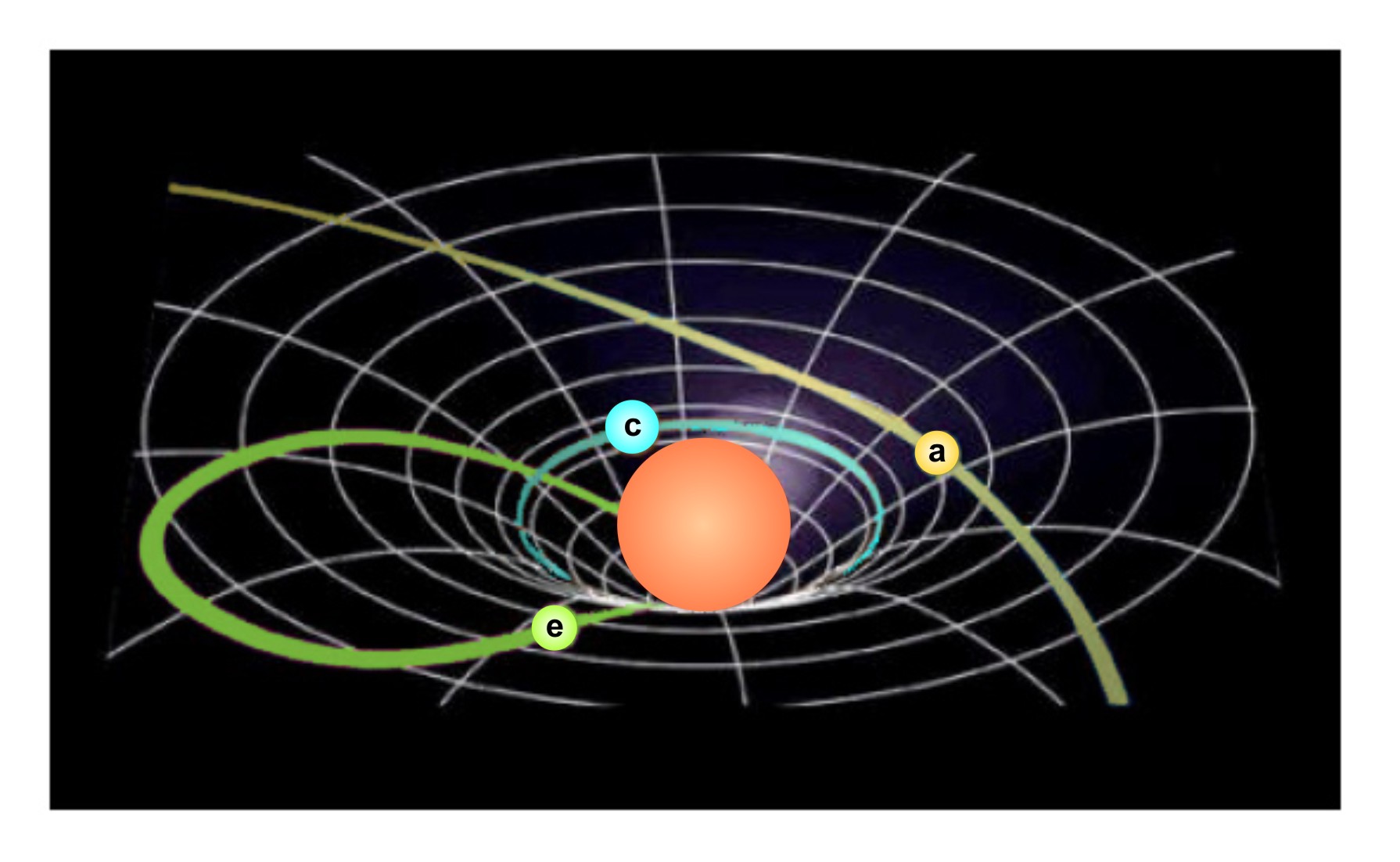}}
\caption{\it Pictorial illustration of a hyperbolic open orbit `a', and of two closed orbits, an elliptic orbit `e' and a circular orbit `c'.}
\label{fig1}
\end{center}
\end{figure}
%%%%%%%%%%%%%%%%%%%

%%%%%%%%%%%%%%%%%%%%%%%%%%%%%
\section*{Teleparallelism: here comes the torsion}
%%%%%%%%%%%%%%%%%%%%%%%%%%%%%

The three-dimensional space we live in~---~as well as the four-dimensional spacetime, in which time plays the role of the fourth dimension~---~has two ever-present fundamental properties: curvature and torsion. In the building up of general relativity, however, Einstein deliberately chose a particular case of spacetime, which may have non-vanishing curvature but has always  vanishing torsion. Such spacetime is known as a Levi-Civita spacetime, after the Italian mathematician Tullio Levi-Civita (1873-1941), a pioneer in the study of such geometries. In this theory, therefore, gravitation is represented solely by curvature: torsion is always kept equal to zero and does not play any role.

On the other hand, it is perfectly conceivable to construct a space with vanishing curvature but non-vanishing torsion. Such space is known as Weitzenb\"ock space~---~after the Austrian mathematician Roland Weitzenb\"ock (1885-1955) who, along with the French mathematician \'Elie Cartan (1869-1951), was a pioneer in the study of spaces with torsion. Similarly to the way a space with curvature only yields general relativity, a space with torsion only yields a completely different gravitational theory, the so-called teleparallel gravity, or simply teleparallelism. In this theory, of course, the gravitational field is represented by torsion, not by curvature.

The mathematical structure behind a space bearing only torsion, known as teleparallel structure, was known since the nineteen twenties and was used by Einstein in his unsuccessful attempt to unify gravitation with electromagnetism. The birth of teleparallel gravity, however, took place a few decades later with the works of the Danish chemist and physicist Christian M{\o}ller (1904-1980). These works had no longer unification purposes~---~were intended exclusively to describe the gravitational interaction. Afterwards, teleparallel gravity received many contributions from different authors, and by now its foundations can be considered to be fully settled. %The authors of the present text have worked for almost two decades in this development, and in 2012 published the first book entirely dedicated to it. 
In what follows some of its outstanding properties will be presented, and possible consequences discussed.

%%%%%%%%%%%%%%%%%%%%%
\section*{Fundamental interactions}
%%%%%%%%%%%%%%%%%%%%%

Three of the four known fundamental interactions of nature~---~namely, the electromagnetic, the weak and the strong interactions~---~are described by a very special kind of theory, known as gauge theory. Only the gravitational interaction, as described by general relativity, does not fit in such a gauge scheme. Teleparallel gravity, on the other hand, fits perfectly in the gauge template. Its advent, therefore, means that now all four fundamental interactions of nature turn out to be described by one and the same kind of theory. Still more amazing is the fact that, although it is a gauge theory~---~and consequently deeply different from general relativity from the conceptual point of view~---~teleparallel gravity is fully equivalent to general relativity in what concerns practical results. Such equivalence shows that gravitation has two equivalent descriptions, and that curvature and torsion are alternative forms of describing the very same gravitational field.

%%%%%%%%%%%%%%%%%%%%%
\section*{Gravitation and inertia}
%%%%%%%%%%%%%%%%%%%%%

Inertial effects were known since Galilei times. They show up whenever the frame from which one observes a phenomenon is accelerated. Consider, for example, the frame represented by a car. When it makes a curve (say) to the left, the passengers feel a force that drive them towards the right. This force is an example of inertial effect. In the geometrical approach of general relativity, gravitation and inertial effects are mixed in a single variable and it is not possible to separate them. In teleparallel gravity, on the other hand, gravitation and inertia are described by different variables, and consequently they can be separated. This is one of the main characteristics of the theory, which has important consequences. In what follows we give two examples of such consequences.

%%%%%%%%%%%%%%%%%%%%%
\section*{Energy of the gravitational field}
%%%%%%%%%%%%%%%%%%%%%

All fundamental fields of nature, as for example the electromagnetic field, have a well--defined energy. One would then expect that the same should happen to the gravitational field. However, it is well-known that, in the context of general relativity, it is not possible to unambiguously define an energy for that field. The reason for this is that in this theory gravitation and inertia are irreversibly mixed, in such a way that whenever one attempts to compute the energy of the gravitational field, the contribution from the inertial effects is automatically included. Since the latter is not well-defined by its very nature~---~it depends on the frame used to describe it~---~the whole energy will not be unambiguously defined either. In the context of teleparallel gravity, however, where gravitation and inertia can be separated, it is possible to obtain independent expressions for the energy of each part. The expression for the energy of the gravitation field alone~---~without the contribution from inertial effects~---~is a well-defined quantity. We can then say that, although equivalent to general relativity, teleparallel gravity is able to provide a solution for an age old problem of gravitation, which is to define unambiguously an energy for the gravitational field.

%%%%%%%%%%%%%%%%%%%%%
\section*{Dispensing with the equivalence principle}
%%%%%%%%%%%%%%%%%%%%%

As is well-known, general relativity is strongly grounded on the weak equivalence principle~---~that is to say, on universality~---~and breaks down in its absence. On the other hand, although it can comply with universality, teleparallel gravity, similarly to Newtonian gravity, does not require the validity of the weak equivalence principle to describe the gravitational interaction, remaining a consistent theory in its absence. One can then say that teleparallelism is a more robust theory in the sense that its existence does not depend on the equality between inertial and gravitational masses. Such characteristic becomes even more important if we remember that there are compelling evidences that universality fails to be true at the microscopic level of quantum mechanics. Since teleparallel gravity remains a consistent theory in its absence, it can be considered a more appropriate formalism to deal with gravitation at the quantum level.

%%%%%%%%%%%%%%%%%%%%%
\section*{A new way to think the gravitational interaction}
%%%%%%%%%%%%%%%%%%%%%

Although equivalent to general relativity, teleparallel gravity introduces new concepts and breaks old paradigms of gravitation. For example, according to general relativity, the presence of a gravitational field produces a curvature in space. As a consequence, the universe itself should be curved. However, from the teleparallel point of view, the attribution of curvature or torsion to spacetime turns out to be a matter of convention as it depends on the theory used to describe gravitation. Another relevant point is that in teleparallel gravity the particle trajectories are not determined by geodesics, but by force equations with torsion playing the role of force. This is similar to the electromagnetic interaction, which is also described by a force. Taking into account the equivalence between general relativity and teleparallel gravity, one can then say that torsion has already been detected: it is responsible for all known gravitational phenomena~---~including the physics of the Solar system~---~as all of them can be reinterpreted in terms of torsion. As a matter of fact, all gravitational phenomena acquire a new perspective when seen from the teleparallel point of view. One can then say that teleparallel gravity is not just an alternative theory, but a whole new way to contemplate the gravitational interaction.

%%%%%%%%%%%%%%%%%%%%%%%%%%%
\section*{Further Readings} % 

\noindent
R. Aldrovandi and J. G. Pereira, {\it Teleparallel Gravity: An Introduction}
(Springer, Dordrecht, 2012).

\noindent
J. G. Pereira, {\it Telaparallelism: a new insight into gravity}, in {\it Springer Handbook of Spacetime}, edited by A. Ashtekar and V. Petkov (Springer, Dordrecht, 2014), 
arXiv:1302.6983.

\end{document}